\documentclass{PoS}

\title{Disentangling singularities with non-linear mappings - a new method for NNLO}

\ShortTitle{Disentangling Singularities with Non-linear Mappings}

\author{\speaker{Franz Herzog}%
         \\
        ETH Zurich\\
        E-mail: \email{fherzog@phys.ethz.ch}}

\abstract{We present a systematic method to factorise singularities using non-linear mappings \cite{FOC}. 
As a first application of the method the fully differential NNLO calculation of the partial decay width of a Higgs boson into 
Bottom-quarks is presented \cite{HBB}.}

\FullConference{10th International Symposium on Radiative Corrections (Applications of Quantum Field Theory to Phenomenology)\\
		 September 26-30, 2011\\
		 Mamallapuram, India}

\begin{document}

\section{Introduction}
With more and more data being collected at the Large Hadron Collider (LHC), 
it is becoming increasingly more important to have control over sizeable higher order QCD corrections 
to the production rates of standard model particles. If theory and experiment are to be compared 
in any meaningful way then these corrections must be computed for arbitrary infrared safe observables 
and cuts on the final state momenta. \\
One major difficulty in tackling fully differential calculations beyond the next-to-leading order (NLO) is 
to handle the intricate infra-red divergences in an efficient way. In this talk we show how non-linear mappings 
can be used to factorise such divergences \cite{FOC}.\\ 
As a proof of concept we report on the fully differential computation of the $H\rightarrow b\bar b$ 
decay width through next-to-next-to-leading-order(NNLO) in the strong coupling expansion \cite{HBB}. 

\section{Observables at NNLO in QCD}
For a given observable $\langle O\rangle$ which may typically be a cross-section 
or a decay rate, we may expand in the strong coupling constant
\begin{equation}
\langle O\rangle=\langle O^{LO}\rangle+\left(\frac{\alpha_s}{\pi}\right)\langle O^{NLO}\rangle
+\left(\frac{\alpha_s}{\pi}\right)^2\langle O^{NNLO}\rangle+\mathcal{O}\left(\alpha_s^3\right)
\end{equation}
We will here focus on the NNLO correction $\langle O^{NNLO}\rangle$, this receives three separate contributions:
\begin{equation}
\langle O^{NNLO}\rangle=\langle O^{RR}\rangle+\langle O^{RV}\rangle+\langle O^{VV}\rangle
\end{equation}
Physically the double-real ($RR$) piece corresponds to the emission of two extra jets in the final state,
the real-virtual ($RV$) corresponds to the one-loop corrections as well as one extra jet in the final state
and the double-virtual ($VV$) corresponds to two-loop corrections. To illustrate these we depict 
interferences of Feynman diagrams corresponding to NNLO corrections to the Higgs to Bottom quark- anti-quark 
decay rate in Figure \ref{fig:H2bb}. Since the final state phase-space for the $RR$, $RV$ and $VV$ is different,
their phase space integrations have to be done separately. The problem which then arises is that all three 
contributions have divergences which only cancel after summing the three contributions. These divergences are conveniently 
regulated in dimensional regularisation, i.e. by setting the dimension to $d=4-2\epsilon$, and the problem is thus reduced
to finding the Laurent expansions of the different NNLO pieces such that after summing them only the finite pieces
remain. However due to the complicated structure of overlapping divergences at NNLO, this a very non-trivial task.
Powerful analytical methods exist to deal with loop integrals \cite{Gehrmann:1999as,Bern:1993kr,Kotikov:1990kg,Smirnov:1999gc,Tausk:1999vh,Anastasiou:2005cb,Czakon:2005rk} 
and can also be applied to fully inclusive phase-space integrals. 
On the contrary dealing with arbitrary final state phase-space cuts requires a purely numerical treatment. 
There are basically two ways in which this can be accomplished:

\begin{figure}
\centering
\includegraphics[scale=1]{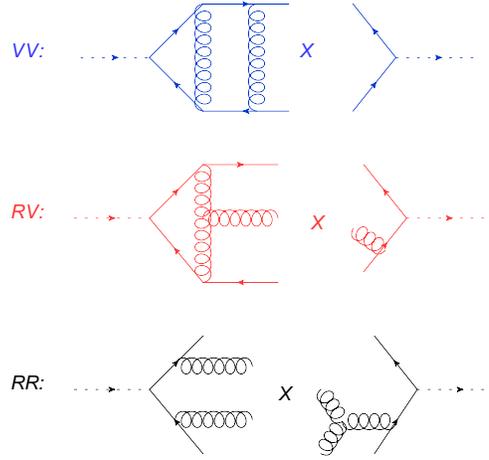}
\caption{Sample interferences of Feynman diagrams contributing to the $H\rightarrow b \bar b$ decay width}
\label{fig:H2bb}
\end{figure}

\begin{enumerate}
\item[i)]  \textbf{The subtraction method}: \\
This method is usually based on QCD factorisation to find counter-terms which reproduce the limits of amplitudes in the different singular kinematic configurations. Since these counter-terms live in kinematic configurations where the concept of 
infrared-safety does not allow for experimental cuts to be imposed, they can be integrated analytically and 
added back to the purely virtual piece, such that poles are cancelled analytically. This method has been applied successfully
in for example \cite{GehrmannDeRidder:2007hr,Weinzierl:2008iv,Catani:2007vq,Catani:2011qz,Ferrera:2011bk}.

\item[ii)] \textbf{The direct integration method}:\\
This method is more brute force and aims at numerically integrating the real pieces directly. 
One way in which this method has been realized successfully is to use \textit{Sector Decomposition} \cite{Binoth:2000ps,Hepp:1966eg,Roth:1996pd,Carter:2010hi} to first factorise the 
singularities and then expand the factorised singularities in plus distributions. This may be easily achieved
using the following identity:
\begin{equation}
x^{-1+a\epsilon}= \frac{\delta(x)}{a \epsilon} + \sum_{n=0}^\infty \frac{(a\epsilon)^n}{n!} \left[\frac{\log^n(x)}{x}\right]_+
\label{eq:plus}
\end{equation}
where $\delta(x)$ is the Dirac-delta function and the plus-distribution is defined such that
$$
\int_0^1 dx f(x)\left[ \frac{g(x)}{x}\right]_+=\int_0^1 dx g(x)\left[ \frac{f(x)-f(0)}{x}\right].
$$
Another way to factorise such overlapping singularities is to use \textit{Non-linear Mappings} \cite{FOC}, 
which we will discuss in more detail in the next section.   
\end{enumerate}

\section{Factorizing overlapping singularities}

\subsection{Sector Decomposition}
We illustrate the method on a simple example integral 
\begin{equation}
I_1= \int_0^1 dx dy \frac{x^{-1+\epsilon}}{x+y} .
\label{eq:I_1}
\end{equation}
To factorise the overlapping singularity at $x=0=y$, we sector decompose by inserting
\begin{equation}
1=\theta(x-y)+\theta(y-x)
\end{equation}
where $\theta(x)$ is the Heaviside step function. In each of the two sectors which are created through this decomposition 
we rescale the integration variable back on the unit hypercube, using $y\mapsto xy$ in the first and $x\mapsto xy$ in the second sector.
This factorizes the integrand since both variables contributing to the overlapping singularity now scale with either $x$ or $y$.
We obtain
\begin{equation}
I_1= \int_0^1 dx dy \frac{x^{-1+\epsilon}}{1+y} +\int_0^1 dx dy \frac{(xy)^{-1+\epsilon}}{1+x} .
\end{equation}
While this procedure is guaranteed to work for any kind of non-factorised singularity, 
it will unavoidably lead to a proliferation of integrals.

\subsection{The non-linear mapping}
A single variable mapping which can be used to disentangle overlapping singularities is
\begin{equation}
x \mapsto \beta(x,y),\quad \beta(x,y) :=  \frac{xy}{(1-x) +y},\quad y>0
\label{eq:beta}
\end{equation}
which keeps the integration boundaries between $0$ and $1$. This mapping allows, for example, to transform the integral $I_1$ into
\begin{equation}
I_1=\int_0^1 dx dy {(xy)}^{-1+{\epsilon}} \left( 1-x+y \right)^{-{\epsilon}},  
\end{equation}
thereby factorizing the singularity at $x=0=y$. It should be noted however that applying $y \mapsto \beta(y,x)$ yields
\begin{equation}
I_1=\int_0^1 dx dy \frac{x^{-1+\epsilon}}{x+(1-y)} 
\end{equation}
which does not factorize the singularity but pushes it to the point $x=0,y=1$. 
This is because the denominator of the Jacobian
\begin{equation}
\frac{d \beta(x,y)}{dx}=\frac{y(1+x)}{[(1-x)+y]^2}
\end{equation}
is not completely cancelled. Indeed there is a general message to be learned here. 
First one should notice that $x$ already constituted a logarithmic singularity stemming from the factor
$x^{-1+\epsilon}$, while $y$ did not since the factor $\frac{1}{x+y}$ is by itself integrable. \\
In \cite{FOC} we therefore termed $x$ to be an \textit{active} and $y$ a \textit{passive} singularity. 
It was then further demonstrated that with the successive use of this transformation 
one can factorize a wide class of different overlapping structures in multidimensional integrals, 
including those which typically occur in NNLO QCD corrections. 
In particular it was demonstrated that all double real corrections in the hadronic production of 
arbitrary massive final states and the most singular two-loop integrals could be factorised with this mapping. 
In \cite{HBB} the mapping was applied to deal with double real and real virtual corrections
to the process $H\rightarrow b\bar b$. We will discuss some details of this calculation in section \ref{HBB}.

\subsection{A two-variable non-linear mapping}
We also mention here a two-variable non-linear transformation \cite{Blumlein:2011wx}
\begin{equation}
(x,y)\mapsto \gamma(x,y),\quad \gamma(x,y) = \left(xy,\frac{x(1-y)}{1-xy}\right)
\label{eq:gamma}
\end{equation}
which was suggested to us by Johannes Bl\"umlein during the RADCOR 2011 conference. 

Interestingly this mapping can be derived from the following one-variable mapping 
\begin{equation}
x\mapsto\alpha(x,A,B),\quad \alpha(x,A,B):=\frac{xA}{(1-x)B+xA},
\end{equation}
which we also employed heavily in \cite{FOC} and \cite{HBB}.
The connection can be seen by noting that
\begin{equation}
\gamma(x,y)=\left(\alpha(x,\alpha(y,1-x,1),1),\alpha(y,1-x,1)\right)=\left(xy,\frac{x(1-y)}{1-xy}\right).
\end{equation}
in other words $(x,y)\rightarrow\gamma(x,y)$ corresponds to the composition of two one-variable mappings,
$x\mapsto\alpha(x,y,1)$ and then $y\mapsto\alpha(y,1-x,1)$.

\section{NNLO corrections to $H\rightarrow b\bar b$}
\label{HBB}
All matrix elements which were required for this calculation were generated with QGRAF \cite{Nogueira:1991ex}. 
Further symbolic manipulations, such as color and Dirac algebra,  
were done with FORM \cite{Vermaseren:2000nd} and MAPLE~\cite{maple}. 
For the computation of loop amplitudes we used the Laporta Algorithm \cite{Laporta:2001dd} 
implemented in AIR \cite{Anastasiou:2004vj} to reduce the amplitude to known master integrals. 
Numerical integrations were done using the VEGAS implementation of the CUBA library \cite{Hahn:2004fe}.

\subsection{Real virtual contribution} 
In the real virtual contribution we encounter integrals of type
$$
\int d\Phi_3 \frac{{}_2F_1(1;-\epsilon,1-\epsilon,-\frac{t}{u} )}{tu} ,
$$
where ${}_2F_1(1;-\epsilon,1-\epsilon,x)$ is Gauss's hypergeometric function, which derives from the 
all orders in $\epsilon$ expansion of the one external mass otherwise massless box. 
Using a phase-space parametrization where the Mandelstam variables, $u$ and $t$, are factorised  
it is suggestive to simply expand them with the plus-distribution eq.(\ref{eq:plus}). 
However the ${}_2F_1(1;-\epsilon,1-\epsilon,x)$ is singular at $x=-\infty$, which is one of the subtraction points. 
This can be seen more clearly when examining the integral representation
\begin{equation}
{}_2F_1(1,-\epsilon,1-\epsilon,-1/z) =z\epsilon \int_0^1 dy \frac{y^{-1-\epsilon}}{z+y}.
\end{equation}
The integral develops an overlapping singularity when $z\rightarrow 0$, similar to our earlier 
example eq.(\ref{eq:I_1}), and can be disentangled with the same mapping.
The mapping simply re-derives the well known "Pfaff's transformation"
\begin{equation}
{}_2F_1(a,b,c;z)=(1-z)^{-b} {}_2F_1(c-a,b,c;\frac{z}{z-1}).
\end{equation}	
Nevertheless the non-linear mapping should lead to more non-trivial results in cases where such identities do not exist.

\subsection{Double real contribution}
One of the most difficult integrals which one faces in the double real is
\begin{equation}
\int d\Phi_4 \frac{J(p_1,p_2,p_3,p_4)}{s_{12}s_{34}s_{14}s_{23}} =\int d\Phi_4 \frac{s_{24}(J(p_1,p_2,p_3,p_4)+J(p_3,p_4,p_1,p_2))}{s_{34}s_{12}s_{23}(s_{13} s_{23}+s_{14} s_{24})}. \label{eq:Iphi4} 
\end{equation}
where $J(p_1,p_2,p_3,p_4)$ is a jet function encoding cuts on the final state momenta, and the 
Lorentz invariants are defined such that $s_{ij}=2p_i.p_j$. 
We use a parametrisation where (we use the shorthand $\bar\lambda_i=1-\lambda_i$)
\begin{eqnarray}
s_{34}  &=&\lambda_1\lambda_2 \nonumber \\
s_{23}  &=&\lambda_1\bar{\lambda}_2 \lambda_4\nonumber \\
s_{24}  &=&\lambda_1\bar{\lambda}_2 \bar{\lambda}_4\nonumber \\
s_{12}  &=&\bar{\lambda}_1\bar{\lambda}_2 \bar{\lambda}_3\nonumber
\end{eqnarray}
 and 
\begin{eqnarray}
s_{13}  &=&\bar{\lambda}_1 \left[ \lambda_4 \lambda_3+\lambda_2\bar{\lambda}_3\bar{\lambda}_4+2\cos(\lambda_5\pi)\sqrt{\lambda_2 \lambda_3 \bar{\lambda}_3 \lambda_4 \bar{\lambda}_4}     \right] \nonumber \\
s_{14}  &=&\bar{\lambda}_1 \left[ \lambda_3 \bar{\lambda}_4+\lambda_2\bar{\lambda}_3\lambda_4-2\cos(\lambda_5\pi)\sqrt{\lambda_2 \lambda_3 \bar{\lambda}_3 \lambda_4 \bar{\lambda}_4}     \right]. \nonumber  
\end{eqnarray}
Having partial fractioned and recombined the two terms in the integrand in eq.(\ref{eq:Iphi4}) to avoid the line singularity,  
we end up with a singularity structure similar to
\begin{equation}
I_2=\int_0^1 dxdydz \frac{(xyz)^\epsilon}{xy(xy+z)}.
\end{equation}
The latter can be factorised by first applying $x\mapsto \beta(x,z)$ and then $(y,z) \mapsto (\beta(y,1-x),\beta(z,1-x))$ \cite{FOC}.

\subsection{Numerical results for the $H\rightarrow b\bar b$ decay rate}
Our numerical result for the inclusive decay rate is
$$
\Gamma_{H\rightarrow b\bar b}^{NNLO}=\Gamma_{H\rightarrow b\bar b}^{LO}\left[ 1 + \left(\frac{\alpha_s}{\pi} \right) 5.6666(4)       + \left(\frac{\alpha_s}{\pi} \right)^2   29.14(2) +\mathcal{O}(\alpha_s^3) \right]  
$$
which compares well with the known analytic result \cite{Baikov:2005rw}
$$
\Gamma_{H\rightarrow b\bar b}^{NNLO}=\Gamma_{H\rightarrow b\bar b}^{LO}\left[ 1 + \left(\frac{\alpha_s}{\pi} \right) 5.6666666..       + \left(\frac{\alpha_s}{\pi} \right)^2   29.146714.. +\mathcal{O}(\alpha_s^3) \right].  
$$
We also present the 2,3 and 4 jet rates using the Jade algorithm \cite{Bartel:1986ua} with $y_{cut}=0.01$:
\begin{eqnarray}
\Gamma_{H\rightarrow b\bar b}^{LO}(4 {\rm Jet Rate})   &=& \Gamma_{H\rightarrow b\bar b}^{LO}\left[  + \left(\frac{\alpha_s}{\pi} \right)^2 94.1(1)    +\mathcal{O}(\alpha_s^3) \right]  \nonumber \\
\Gamma_{H\rightarrow b\bar b}^{NLO}(3 {\rm Jet Rate})  &=& \Gamma_{H\rightarrow b\bar b}^{LO}\left[  + \left(\frac{\alpha_s}{\pi} \right) 19.258(4)       + \left(\frac{\alpha_s}{\pi} \right)^2  241(2)   +\mathcal{O}(\alpha_s^3) \right]   \nonumber \\
\Gamma_{H\rightarrow b\bar b}^{NNLO}(2 {\rm Jet Rate}) &=& \Gamma_{H\rightarrow b\bar b}^{LO}\left[ 1 - \left(\frac{\alpha_s}{\pi} \right)13.591(6)        - \left(\frac{\alpha_s}{\pi} \right)^2 307(2)    +\mathcal{O}(\alpha_s^3) \right]   
\end{eqnarray}

\begin{figure}
\begin{center}
\includegraphics[scale=1]{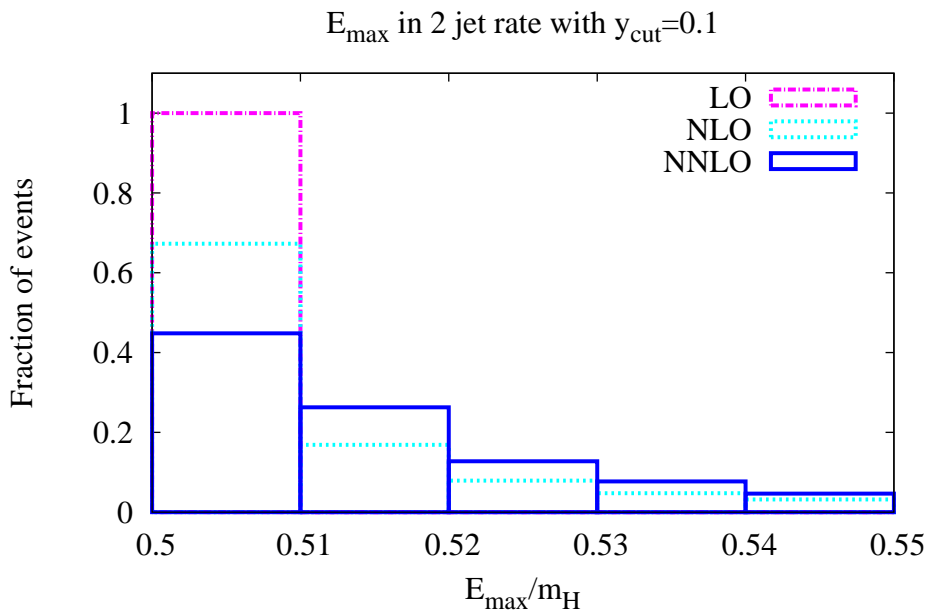}
\caption{The maximum energy of the leading jet in the 2 jet rate computed with JADE using $y_{cut}=0.1$ in the $H\rightarrow b \bar b$ width.}
\label{fig:Emax}
\end{center}
\end{figure}

In Figure \ref{fig:Emax} we also present a fully differential observable.

\section{Conclusion}
In these proceedings we presented a method to factorize singularities at NNLO using a single non-linear mapping \cite{FOC}.
We showed that the method can be used as a direct integration method and thus can be seen as an alternative to Sector decomposition,
without proliferating the number of integrals. We demonstrated that this method could be used to do entire NNLO 
calculations by presenting the complete fully differential calculation of the $H\rightarrow b \bar b$ decay width \cite{HBB}. 
We find that our inclusive result is in good agreement with the known analytical result and further present 
$2, 3$- and $4$ jet rates with the JADE algorithm as well as the distribution of the maximum energy of the leading jet in the 
$2$-jet rate.

\section*{Acknowledgements}
We thank the organizers of RADCOR 2011 for an excellent workshop. 
We also thank Johannes Bl\"uhmlein and Gudrun Heinrich for interesting discussions.
This research is supported by the ERC Starting Grant for the project
``IterQCD'' and the Swiss National Foundation under contract SNF
200020-126632.

\end{document}